\documentclass[aps,preprintnumbers,superscriptaddress,twocolumn,amsmath,amssymb,floatfix,prb]{revtex4} 
\pdfoutput=1
\usepackage{graphicx}
\usepackage[usenames,dvipsnames]{xcolor}
\usepackage[colorlinks,bookmarks=false,citecolor=NavyBlue,linkcolor=Red,urlcolor=blue]{hyperref}
\usepackage{dsfont}
\usepackage[export]{adjustbox}
\usepackage{bbold}

\newcommand\be{\begin{equation}}
\newcommand\bea{\begin{eqnarray}}
\newcommand\bes{\begin{subequations}}
\newcommand\esu{\end{subequations}}
\newcommand\ee{\end{equation}}
\newcommand\eea{\end{eqnarray}}

\newcommand{\cmmnt}[1]{}

\def\doi{http://dx.doi.org/}

\usepackage{braket}

\newcommand\ocite[1]{[\onlinecite{#1}]}

\begin{document}

\title{Superluminal moving defects in the Ising spin chain
}

\author{Alvise Bastianello}
\affiliation{SISSA \& INFN, via Bonomea 265, 34136 Trieste, Italy}

\author{Andrea De Luca}
\affiliation{The Rudolf Peierls Centre for Theoretical Physics, Oxford University, Oxford, OX1 3NP, United Kingdom}
%\date{\today}

\begin{abstract}
Quantum excitations in lattice systems always propagate at a finite maximum velocity. We probe this mechanism by considering a defect travelling at a constant velocity in the quantum Ising spin chain in transverse field. Independently of the microscopic details of the defect, we characterize the expectation value of local observables at large times and large distances from the impurity, where a Local Quasi Stationary State (LQSS) emerges. The LQSS is strongly affected by the defect velocity: for superluminal defects, it exhibits a growing region where translational invariance is spontaneously restored. We also analyze the behavior of the friction force exerted by the many-body system on the moving defect, which reflects the energy required by the LQSS formation. Exact results are provided in the two limits of extremely narrow and very smooth impurity. Possible extensions to more general free-fermion models and interacting systems are discussed.
\end{abstract}


\maketitle
%\tableofcontents

\section{Introduction}

Recent times have witnessed an increasing interest in the out-of-equilibrium dynamics of isolated quantum many-body systems, sustained by various experimental achievements \ocite{exp1,exp2,exp3,exp4,exp5,exp6,exp7,exp8,exp9,exp10,exp11,exp12,exp13,exp14,exp15}.
Several efforts have been devoted to the understanding of \emph{quantum quenches} \ocite{calabrese-cardy}, in which a closed quantum system is suddenly perturbed and then let evolve: despite the \emph{unitary} evolution of the whole many-body system, the expectation values of local observables can still exhibit relaxation \ocite{pssv}.
With a global quench, an extensive amount of energy is injected in the system and the late-time relaxation only depends on those quantities which are conserved by the dynamics. In this perspective, quantum systems can be broadly divided into two classes: in generic systems, few local operators (e.g. Hamiltonian) are conserved and the system is expected to relax towards a standard Gibbs Ensemble. Instead, in one dimension, there exist integrable models \ocite{Korepin,smirnov,takahashi}, which possess infinitely many local conserved quantities that constrain the dynamics.
Homogeneous quantum quenches in integrable models have been tackled in an impressive number of works \ocite{ggew1,ggew2,ggew3,ggew4,ggew5,ggew6,ggew7,ggew8,ggew9,ggew10,specialissue} and are nowadays well-understood in terms of the so-called Generalized Gibbs Ensemble (GGE) \ocite{gge1}, constructed out of a complete set of local and quasilocal integrals of motion \ocite{ggef1,ggef2,ggef3,ggef4, lch1,lch2,lch3,lch4,lch5,lch6,lchf1,lchf2,lchf3,lchf4,lchf5}.

The experience gained in the homogeneous framework proved to be crucial in approaching \emph{inhomogeneous} initial states. A particularly simple example is the \textit{partitioning} protocol, first introduced in the framework of conformal field theories~\ocite{NESSc1,NESSc2,NESSc3,NESSc4,NESSc5,NESSc6}. Several works including
 numerical investigations \ocite{NESSnum1,NESSnum2,NESSnum3}, results in free models  \ocite{NESSf0,NESSf1,NESSf2,NESSf3,NESSf4,NESSf5,NESSf6,NESSf7,NESSf8,NESSf9,NESSf10,NESSf11,NESSf12} and approximate predictions~\ocite{NESSnum2,NESSI2}
finally led to the formulation of the Generalized Hydrodynamics (GHD)~\ocite{GHD1, GHD2}, holding for a large class of integrable systems.
In practice, because of integrability, the dynamics can be understood as the elastic scattering of stable quasiparticles: this leads to a ballistic scaling, so that, at large times the system can be thought to be locally equilibrated to a GGE. In particular, finite distances from the junction ($x/t \to 0$) reach Non-Equilibrium Steady States \ocite{NESSI2,NESSI3}, while a Locally Quasi-Stationary State (LQSS) \ocite{GHD1,GHD2} describes the behavior at fixed $x/t \equiv \zeta \neq 0$. 
This approach has rapidly lead to a plethora of accurate results in a wide range of situations~\ocite{GHD1,GHD2,GHD3,GHD6,GHD7,GHD8,GHD10,F17,DS,ID117,DDKY17,DSY17,DS2,ID217,CDV17,dcorr,BDWY17, mazza2018}.

A similar phenomenology is expected for defect protocols (or local quenches), where a localized impurity is suddenly activated in an otherwise homogeneous system, even though exact results are only known for free theories~\ocite{localquenches1,localquenches2,localquenches3,localquenches4}. Remarkably, a non-trivial LQSS can emerge despite the extreme locality of the perturbation.

As long as lattice systems are concerned,
all these settings possess a finite maximal velocity $v_M < \infty$ for the propagation of quasiparticles, being it fundamentally due to the Lieb-Robinson bound \ocite{liebrob}. 
The existence of $v_M$ makes natural to wonder what happens in the case of a defect \emph{moving} at constant velocity $v$. In particular, can the system cope with the defect when the latter moves faster than $v_M$? Is an LQSS still formed and how is it affected by the velocity of the defect?
The velocity $v_M$ has the same role of the speed of light in the spreading of information, therefore for the sake of convenience, we refer to fast defects $v>v_M$ as ``superluminal", while defect moving slower than $v_M$ will be called ``subluminal".

Besides being a compelling question from a theoretical point of view, similar out-of-equilibrium protocols already underwent experimental investigation (see for example Ref. \ocite{exp15}).
 
The best candidates to acquire some physical insight in this problem are  \emph{free models}. 
While moving impurities have been considered in different contexts (see for example Ref.\ocite{refB1,refB2,refB3,refB4,AsPi04,BuGe01}), the investigation of the LQSS was only recently presented in Ref. \ocite{hopdef} in a system of hopping fermions.
In this case, a superluminal moving defect $v>v_M$ has been proven to not affect the system, preventing any LQSS formation and thus establishing a clear threshold $v=v_M$.
A crucial feature of the case considered in \ocite{hopdef} was the presence of a $U(1)$ symmetry, associated with the conservation of the number of fermions. Thus, the defect acted as a moving scattering potential, but could not create or destroy quasiparticles. As a consequence, in a non interacting system possessing $U(1)$ symmetry (i.e. particle-number conservation) such as Ref. \ocite{hopdef}, a superluminal defect could not experience any friction force. 

\begin{figure}[t!]
\includegraphics[width=0.95\columnwidth]{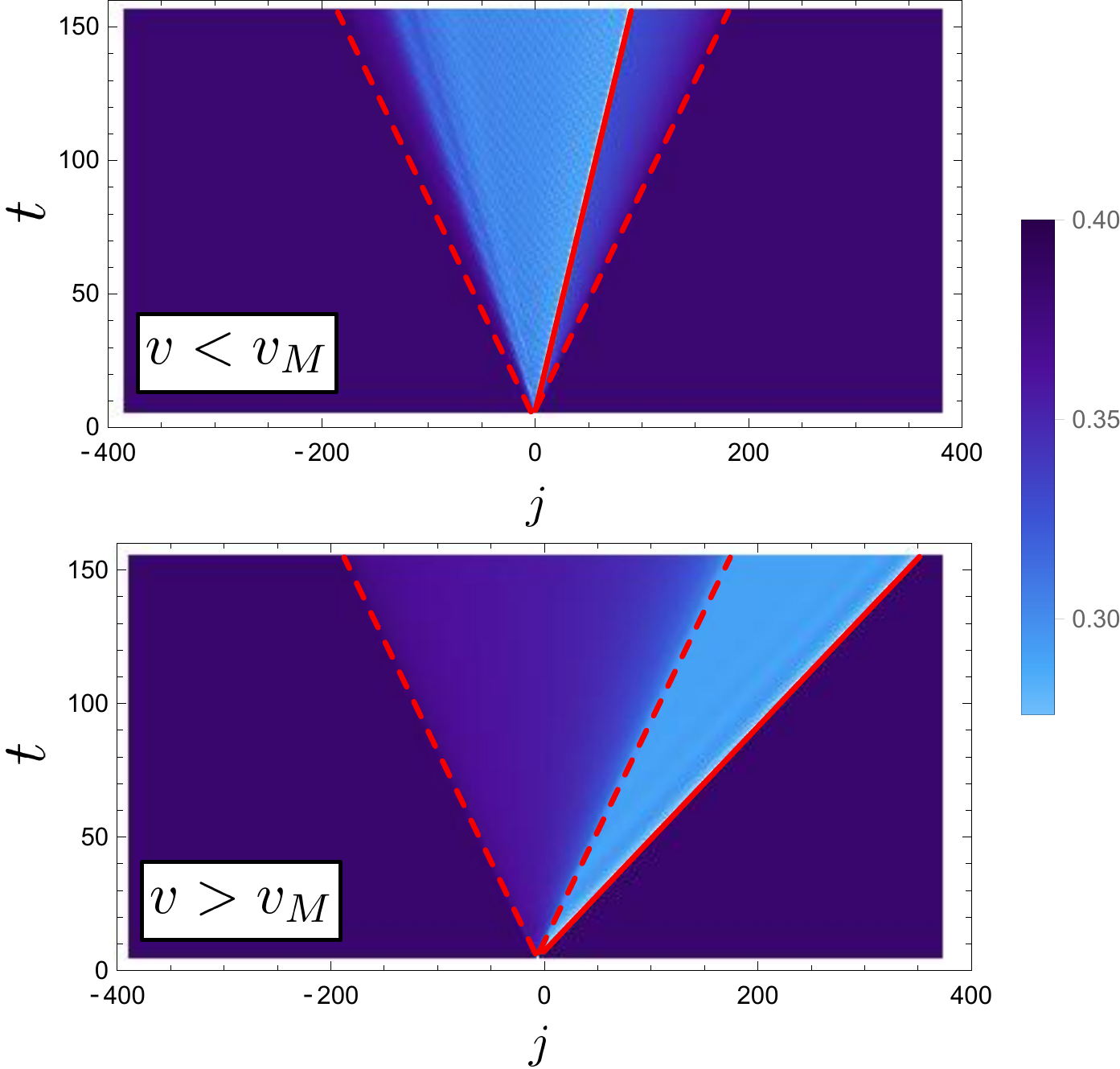}
\caption{\label{LQSSligthcone} Spreading of the LQSS due to a moving defect (continuous red line), for a subluminal $(v<v_M)$ and superluminal ($v>v_M$) defects. In both cases, the local magnetization density $\langle \hat{\sigma}^z\rangle$ is considered.
In the subluminal case, the perturbation is entirely contained within the light-cone spreading from the initial position of the defect (dashed lines), while in the superluminal case the system is affected beyond such a lightcone, following the defect.
In both cases, $v_M=1$ and a $\delta-$like defect is considered (see Section \ref{deltaSolution} for notation).
Parameters used for $v<v_M$: $h=1.15$, $c=0.5$, $v=0.5$.
Parameters used for $v>v_M$: $h=1.15$, $c=2$, $v=2$.
}
\end{figure}

In this work we will address the same set up in more general frameworks, where the particle number is not conserved. This purpose leads us to consider the paradigmatic example of the Ising spin $1/2$ chain in a transverse field, however both the tools and the resulting phenomenology hold for all quadratic fermionic models, completing the puzzle of which Ref. \ocite{hopdef} was a first piece. 
The Ising model is defined through the following Hamiltonian
\be
\hat{H}=-\frac{1}{2}\sum_{j=1}^N \hat{\sigma}_j^x\hat{\sigma}_{j+1}^x+h\hat{\sigma}_j^z\, ,\label{Hising}
\ee
where $\hat{\sigma}_j^{x,y,z}$ are the standard Pauli matrices, periodic boundary conditions are assumed even though we will work directly in the thermodynamic limit. 
At time $t=0$, we activate a moving impurity in the transverse magnetic field
\be
h\to h(j-vt)=h+\delta h(j-vt)\label{defect_activation}\, .
\ee
By means of a Jordan-Wigner transformation, Eq.~\eqref{Hising} can be recast into a system of non-interacting spinless fermions on a lattice. We anticipate that for superluminal defects, particles are turned into holes after scattering on the defect. In this way, the LQSS appears even for $v>v_M$ (see Fig. \ref{LQSSligthcone}), an effect forbidden by the $U(1)$-symmetry analyzed in Ref. \ocite{hopdef}.
The behavior of the superluminal LQSS is nevertheless 
very different from the $v<v_M$ case. We will show that a superluminal LQSS is a purely \emph{quantum} effect: for very smooth and broad defects the semiclassical approximation is well-justified and and in this limit the superluminal LQSS disappears.

The paper is so organized: in Section \ref{secGenIsing} we provide a brief summary of the homogeneous Ising model and set up the notation. In Section \ref{secScattering}, the problem of a generic moving defect is framed in a convenient scattering theory, with some details left to Appendix \ref{appScat}. The produced LQSS is expressed only in terms of the scattering of quasiparticles on the defect and discussed in generality.
In Section \ref{deltaSolution}, we apply the scattering method to the limiting case of an extremely narrow defect, namely a $\delta-$function. The results are then tested against the numeric.
In Section \ref{secSemiclassic}, the opposite limiting case of an extremely smooth defect is considered and solved within a semiclassical theory, which derivation is contained in Appendix \ref{derivation_semiclassical}. As anticipated, a supersonic LQSS is absent in this case, clarifying that the LQSS for superluminal defects is a purely quantum effect.
Section \ref{frictionsec} is devoted to a concrete and experimentally relevant manifestation of the LQSS: the friction exerted by the medium on the defect. We discuss its behavior for different defect velocities (see also Appendix \ref{app_fastdef}).
Finally, Section \ref{secConclusions} gathers our conclusions and
Appendix \ref{numApp} describes the numerical methods we employed.
\section{Generalities of the Ising model}
\label{secGenIsing}
The solution of the homogeneous Ising model is well known \ocite{solIsing} and nowadays it is a cornerstone of several ordinary textbooks of quantum mechanics, nevertheless hereafter we provide an useful short summary to set our notation. The Hamiltonian (\ref{Hising}) can be mapped to a system of free fermions through a Jordan-Wigner transformation, the fermionic operators being defined as
\be
\hat{d}_j=e^{i\pi\sum_{l=1}^{j-1}\hat{\sigma}^+_l\hat{\sigma}_l^-}\hat{\sigma}^+_j\, ,\label{JW}
\ee
where $\hat{\sigma}_l^\pm=(\hat{\sigma}^x_l\pm i\hat{\sigma}^y_l)/2$. We are assuming the lattice sites to run from $1$ to $N$ with periodic boundary conditions, but the thermodynamic limit will always be considered. The operators $\hat{d}_j$ are spinless fermions obeying the anticommutation rules $\{\hat{d}^\dagger_j,\hat{d}_{j'}\}=\delta_{j,j'}$. Despite the non-locality of the Jordan-Wigner transformation, the Ising Hamiltonian is still local in the fermionic basis
\be
\hat{H}=\sum_{j=1}^N-\frac{1}{2}\left(\hat{d}^\dagger_j \hat{d}^\dagger_{j+1}+\hat{d}^\dagger_j\hat{d}_{j+1}+\text{h.c.} \right)+h\hat{d}^\dagger_j\hat{d}_j\, .\label{Hfermion}
\ee
Above,  ``h.c." stands for the hermitian conjugated of the expression.
(Anti)periodic boundary conditions in the fermion basis must be used in the (even)odd magnetization sectors, but such a difference is irrelevant in the thermodynamic limit in which we are ultimately interested.
In the Fourier space a Bogoliubov transformation
\be
\hat{\psi}_j=\begin{pmatrix}\hat{d}_j \\ \hat{d}^\dagger_j \end{pmatrix}=\int_{-\pi}^\pi \frac{dk}{\sqrt{2\pi}}\, e^{ikj}\begin{pmatrix}\cos\gamma_k && i\sin\gamma_k \\ i\sin\gamma_k && \cos\gamma_k  \end{pmatrix}\begin{pmatrix}\hat{\alpha}_k \\ \hat{\alpha}^\dagger_{-k}\end{pmatrix}\, ,
\ee
diagonalizes the Hamiltonian
\be
\hat{H}=\int_{-\pi}^\pi dk\, \omega(k)\hat{\alpha}^\dagger_k\hat{\alpha}_k+\text{const.}
\ee
The operators $\hat{\alpha}_k$ are canonical Fermionic operators $\{\hat{\alpha}^\dagger_k,\hat{\alpha}_q\}=\delta(k-q)$ and $\omega(k)=\sqrt{(\cos k-h)^2+\sin^2k}$, while the Bogoliubov angle $\gamma_k$ must be chosen in such a way
\be
\tan\gamma_k=\frac{\omega(k)+\cos k-h}{\sin k}\, .\label{bogangle}
\ee

The velocity of the excitations is readily identified with the group velocity 
\be
v(k)=\partial_k\omega(k)
\ee
and it is clearly bounded, leading to a finite $v_M$. In particular, it holds true
\be\label{eqvc}
v_M=\min(h,1)\, .
\ee

The Ground State is easily identified as the vacuum with respect to $\hat{\alpha}_k$, i.e. the state such that $\hat{\alpha}_k|GS\rangle=0$ for every momentum. This is a Gaussian state, i.e. multipoint correlators of $\hat{\alpha}_k$ factorize in two point correlators, unambiguously fixed by the excitation density $\eta(k)\delta(k-q)=\langle \hat{\alpha}^\dagger_k \hat{\alpha}_q\rangle$, being the latter zero in the GS. In the following, we will assume our system to be always initialized in a Gaussian ensemble diagonal in the eigenmodes $\hat{\alpha}_k$, with a possible non-trivial excitation density $\eta(k)$. Among these states and apart from the Ground State, relevant examples are thermal states, and more generally GGE density matrix \ocite{gaussification1,gaussification2,gaussification3,gaussification4}
\be
\rho_\text{GGE}\propto \exp\left[-\int_{-\pi}^\pi dk \log(2\pi\eta^{-1}(k)-1) \hat{\alpha}^\dagger_k \hat{\alpha}_k\right]\, .\label{GGEmatrix}
\ee
These states are translationally invariant and diagonal in the eigenmodes of the Hamiltonian in the absence of the impurity, therefore the non-trivial time evolution will be entirely caused by the defect. Note that the modes $\hat{\alpha}_k$ and the fermions $\hat{d}_j$ are linked by a linear transformation, therefore a Gaussian state in terms of the modes $\hat{\alpha}_k$ translates into a Gaussian state in the real-space field operators. In other words, any correlator of the $d_j$ operators is determined in terms of $\langle\hat{\psi}_j\hat{\psi}^\dagger_{j'} \rangle$, 
which in our notation is a $2\times 2 $ matrix
\be\label{two_point}
\langle \hat{\psi}_j\hat{\psi}^\dagger_{j'}\rangle=\begin{pmatrix} \langle \hat{d}_j\hat{d}^\dagger_{j'}\rangle && \langle \hat{d}_j\hat{d}_{j'}\rangle \\ \langle \hat{d}^\dagger_{j}\hat{d}^\dagger_j\rangle && \langle \hat{d}^\dagger_j\hat{d}_{j'}\rangle\end{pmatrix}\, .
\ee

\section{The moving defect and the scattering problem}
\label{secScattering}

Following the proposed out-of-equilibrium protocol, at time $t=0$, we activate the moving defect Eq. \eqref{defect_activation}. The dynamics is encoded in the Heisenberg equations of motion
\be
i\partial_t\hat{\psi}_j=\frac{(i\sigma^y-\sigma^z)}{2}\hat{\psi}_{j-1}-\frac{(i\sigma^y+\sigma^z)}{2}\hat{\psi}_{j+1}+\sigma^z h(j-vt) \hat{\psi}_{j}\, , \label{EQM}
\ee
whose linearity preserves the gaussianity of the ensemble along the whole time evolution, therefore only the two point correlators \eqref{two_point} are needed to completely describe the state.
Note that we employ the notation of Pauli matrices without the ``hat'' to clarify that they act on the two components of $\hat{\psi}_j$, rather than on the many-body Hilbert space as in Eq.~\eqref{Hising}.
The explicit time dependence in Eq. \eqref{EQM} introduced by the defect prevents a straightforward solution. On the other hand, time-dependent perturbation theory is not suited to study late-time dynamics.

Following Ref. \ocite{hopdef}, our strategy is then to perform a change of reference frame which poses the defect at rest: in this way, any time dependence of the equations of motion is removed.
Of course, the discreteness of the underlying lattice, compared with the continuous shift of the moving defect, obstructs such a transformation. 

In order to overcome this issue, we first map the discrete problem in a continuous one, where such a change of reference system is possible. There, the late-time dynamics can be solved in terms of a scattering theory and the final solution is then pulled back to the original discrete problem.

\subsection{Mapping the discrete in the continuous}
\label{secContinuous}
The formal solution of the equation of motion Eq. (\ref{EQM}) can be expressed in terms of the Green function $G_{j,j'}(t)$, a $2\times 2$ matrix, as
\be
\hat{\psi}_j(t)=\sum_{j'} G_{j,j'}(t)\hat{\psi}_{j'} \;.
\ee
The Green function satisfies the differential equation
\begin{multline}
i\partial_tG_{j,j'}(t)=\frac{(i\sigma^y-\sigma^z)}{2}G_{j-1,j'}(t)\label{dG}\\
-\frac{(i\sigma^y+\sigma^z)}{2}G_{j+1,j'}(t)+\sigma^z h(j-vt) G_{j,j'}(t)\, ,
\end{multline}
together with the initial condition $G_{j,j'}(0)=\delta_{j,j'}\mathbb{1}$. 

The anticipated mapping to a continuous system can be achieved defining a closely-related \emph{continuous} Green function $\mathcal{G}_{x,x'}(t)$. Hereafter, $x$ and $x'$ are continuous real variables and $\mathcal{G}_{x,x'}(t)$ is defined as the unique solution of
\begin{multline}
i\partial_t\mathcal{G}_{x,x'}(t)=\frac{(i\sigma^y-\sigma^z)}{2}\mathcal{G}_{x-1,x'}(t)\label{cG}\\
-\frac{(i\sigma^y+\sigma^z)}{2}\mathcal{G}_{x+1,x'}(t)+\sigma^z h(x-vt) \mathcal{G}_{x,x'}(t)\, ,
\end{multline}
with the initial condition $\mathcal{G}_{x,x'}(0)=\delta(x-x')\mathbb{1}$. 
Given that Eq. \eqref{cG} is nothing else than Eq. \eqref{dG} in which discrete coordinates are promoted to continuous ones, $G$ and $\mathcal{G}$ are not surprisingly related. More specifically, since Eq. (\ref{cG}) couples together only points that differ of an integer distance, we can rewrite 
\be
\mathcal{G}_{x,x'}(t)=2\pi \delta(e^{i2\pi(x-x')}-1)g_{x,x'}(t)\, ,\label{singg}
\ee
and using Eq. \eqref{singg} in Eq. \eqref{cG}, it can checked that  $g_{x,x'}(t)$ for integer $x,x'$ satisfies the same differential equation and initial conditions as $G_{j,j'}(t)$. 
Therefore, we can identify $G_{j,j'}(t)=g_{j,j'}(t)$ and work directly with the continuous variables.

Moreover, being the initial state most easily expressed in the momentum space, it is convenient to take a Fourier transform with respect to the second coordinate. 
Then, one observes that the Dirac $\delta$ in Eq. (\ref{singg}) transforms the Fourier integral in a Fourier series. In practice, we arrive at
\begin{multline}
\tilde{\mathcal{G}}_{j,k}(t)=\int dx \,e^{-ik x} \mathcal{G}_{j,x}(t)=\\ = \sum_{j'} e^{-ikj'}g_{j,j'}(t)=\tilde{G}_{j,k}(t)\label{gfourier}
\end{multline}
where we explicitly used $G_{j,j'}(t)=g_{j,j'}(t)$.
We remark that since the Green function $\mathcal{G}$ is defined on a continuous space, its Fourier components run over the entire real axis. 
On the contrary, the Green function $G$ is defined on a lattice and thus its Fourier transform is naturally embedded in a Brillouin zone. 
However, Eq.~\eqref{singg} makes the l.h.s. of Eq. (\ref{gfourier}) periodic in the momentum space, as it should be for Eq. (\ref{gfourier}) to hold true.

The advantage of dealing with a continuous system is that we can now perform a change of coordinates. Defining $\mathcal{G}^{(v)}_{x,x'}(t)=\mathcal{G}_{x+vt,x'}(t)$, we obtain a time-independent equation of motion
\begin{multline}
\label{ccG}
i\partial_t\mathcal{G}^{(v)}_{x,x'}(t)=iv\partial_x\mathcal{G}^{(v)}_{x,x'}(t)+\frac{(i\sigma^y-\sigma^z)}{2}\mathcal{G}^{(v)}_{x-1,x'}(t)
\\-\frac{(i\sigma^y+\sigma^z)}{2}\mathcal{G}^{(v)}_{x+1,x'}(t)++ \sigma^z h(x) \mathcal{G}^{(v)}_{x,x'}(t) \, ,
\end{multline}
but which now explicitly involves the defect's velocity $v$.
Eq. (\ref{ccG}) can be interpreted as a single-particle Schr\"oedinger equation, which can be solved in terms of its eigenfuctions. 
This spectrum will depend on the specific choice of the function $h(x)$; however in the hypothesis of a narrow defect, the problem is most easily addressed in the framework of a scattering theory.

\subsection{The scattering theory and the LQSS}
\label{secSc}

The equation of motion of the Green function Eq. \eqref{ccG} is most easily solved in terms of the eigenfunctions, i.e.
\be
\mathcal{G}^{(v)}_{x,x'}=\sum_{a=1,2}\int \frac{dk}{2\pi} e^{-iE^a(k) t}\phi_{k,a}(x) \phi^\dagger_{k,a}(x')\, ,\label{Geigen}
\ee
where $\phi_{k,a}$ constitute a complete set of two dimensional vectors satisfying the time-independent version of the Schr\"oedinger equation (\ref{ccG})
\be
E^a(k)\phi_{k,a}(x)=(H_0\phi_{k,a})(x)+\sigma^z\delta h(x)\phi_{k,a}(x)\, .\label{sch}
\ee
Above, in the linear operator $H_0$ we collected the homogeneous part of the equation, while the inhomogeneous term due to the defect is written explicitly. The notation $\phi_{k,a}$ has been chosen in the perspective of the scattering theory we are going to develop. We stress that $\phi_{k,a}(x)$ is a two dimensional vector, not a quantum operator: in this framework $\phi_{k,a}(x)\phi^\dagger_{k,a}(x')$ represents a $2 \times 2$ matrix, similarly to the notation of Eq. \eqref{two_point}. 
The eigenfunctions satisfy the orthonormality and completeness relations
\be
\int dx\, \phi^\dagger_{k,a}(x)\phi_{q,b}(x)=\delta(k-q)\delta_{a,b}\, ,\label{constr1}
\ee
\be
\delta(x-x')=\sum_{a=1,2}\int dk\, \phi^\dagger_{k,a}(x)\phi_{k,a}(x')\, .\label{constr2}
\ee

Taking advantage of the fact that $\delta h(x)$ is a localized impurity, we can describe the $\phi_{k,a}$ eigenfunctions in a scattering theory framework, in particular they must satisfy the Lippmann-Schwinger equation \ocite{sakurai} having as source terms the eigenfunctions of the homogeneous problem. With the definition
\be
u_{k,1}=\begin{pmatrix}\cos\gamma_k \\ i\sin\gamma_k \end{pmatrix}\,, \hspace{1pc} u_{k,2}=\begin{pmatrix}i\sin\gamma_k \\ \cos\gamma_k \end{pmatrix}\,\label{eigenvectors}
\ee
The Lippmann-Schwinger equation is simply
\begin{multline}
\phi_{k,a}(x)=\frac{e^{ikx}}{\sqrt{2\pi}}u_{k,a}+\label{lippman}\\
+\int dx' \left[\frac{1}{E^a(k)-H_0+i0^+}\right]_{x,x'}\delta h(x')\sigma_z \phi_{k,a}(x')\, .
\end{multline}
Above, $e^{ikx}u_{k,a}$ are the eigenfunctions of the homogeneous problem, with energies
\be
E^1(k)=-vk +\omega(k)\, ,\hspace{1pc} E^2(k)=-vk -\omega(k)\, .\label{scaten}
\ee
We can recognize $E^1(k)$ as the energy of excitations  $\omega(k)$ with an additional shifting $-vk$ due to the change of reference frame. Similarly, $E^2(k)$ represents the energy of a hole.  

As we will see, the LQSS  is completely determined by the behaviour of the eigenfunctions in the scattering region, i.e. far away from the defect. From Eq.~(\ref{lippman}), the large distance expansion is immediately recovered from the singularities in the kernel
\begin{multline}
\phi_{k,a}(x)\simeq \frac{1}{\sqrt{2\pi}}\Bigg(\theta(-xv^a(k))e^{ikx}u_{k,a}+\sum_{b}\int dq\, e^{iqx}\times\label{asymp}\\
\delta(E^b(q)-E^a(k))\theta(xv^b(q)) |v^b(q)|S_{b,a}(q,k)u_{q,b}\Bigg)\, ,
\end{multline}
where $\theta(x)$ indicates the Heaviside Step function.
This equation admits a simple interpretation: the step functions $\theta(x)$ in Eq. \eqref{asymp} discern between incoming/outgoing states; the second term describes the scattering of a mode $(k,a)$ into a mode $(q,b)$: this is allowed only if the energy is conserved $E^b(q)=E^a(k)$ and it is mediated by the scattering amplitude $S_{b,a}(q,k)$. We remark that the number of solutions to $E^b(q)=E^a(k)$ is always finite for $v\neq 0$.
\begin{figure*}[t!]
\advance\leftskip-1cm
\includegraphics[width=0.95\textwidth]{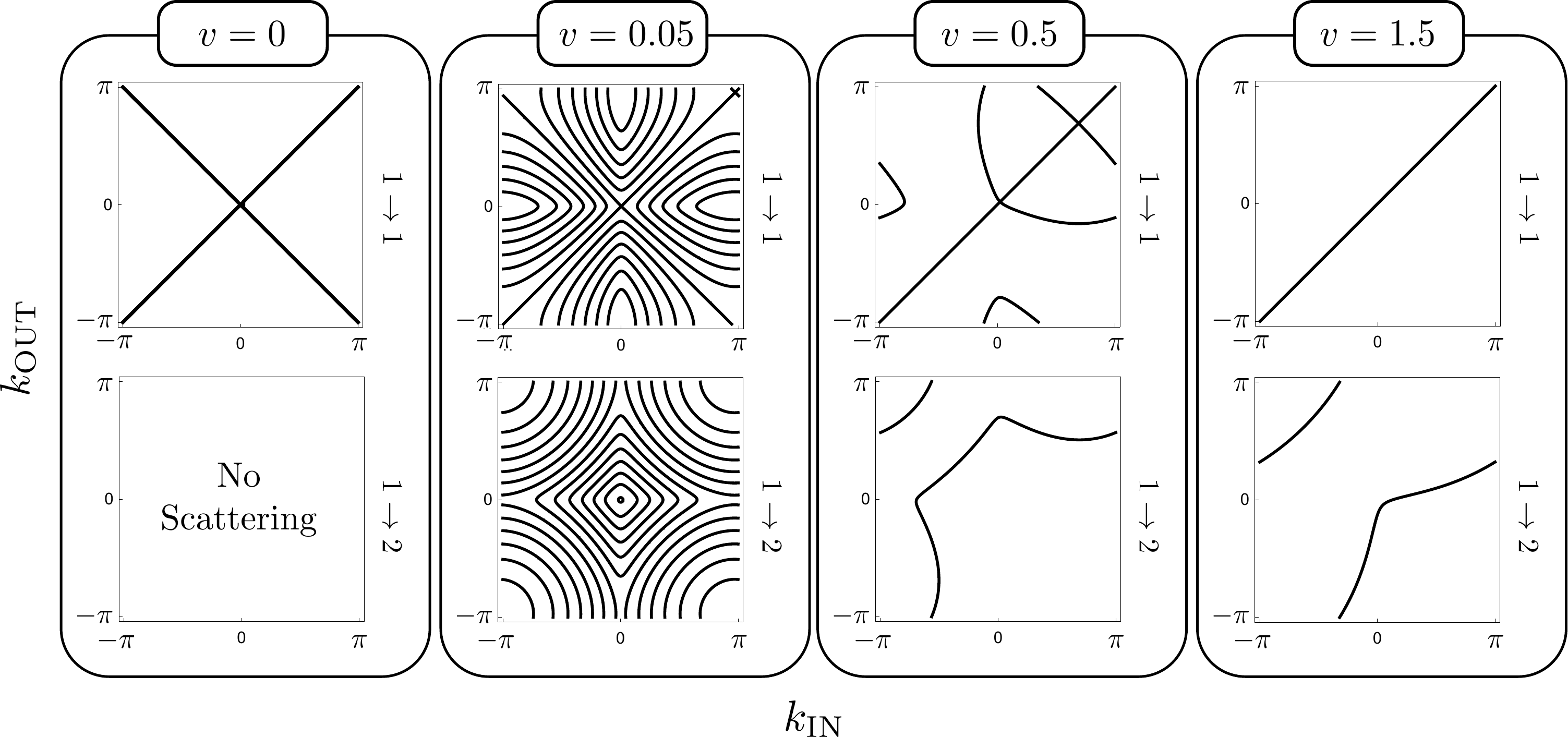}
\caption{\label{figscatchannels} Dynamically accessible scattering channels $1\to1$ and $1\to 2$ (see Section \ref{secSc} for notation). The scattering channels are identified as the solutions of Eq.~\eqref{enercons} which are ultimately pulled back to the first Brillouin zone. The magnetic field is set $h=1.15$, thus $v_M=1$. 
}
\end{figure*}
In Eq. \eqref{asymp}, $v^a(k)$ represents the velocity of the excitation computed in the moving reference frame, defined as $v^a(k)=\partial_k E^a(k)$. In particular, from Eq.~\eqref{asymp}, it follows
\be
v^1(k)=v(k)-v\hspace{2pc} v^2(k)=-v(k)-v\, .\label{vdef}
\ee

The scattering amplitudes can be expressed formally in the form
\begin{multline}
S_{b,a}(q,k)=\\
\begin{cases}1+\int dx\,e^{-ikx}u^\dagger_{k,a}\sigma^z\delta h(x)\phi_{k,a}(x) \,\,\,\, (k,a)=(q,b)\\ \int dx\, e^{-iqx}u^\dagger_{q,b}\sigma^z\delta h(x)\phi_{k,a}(x)   \hspace{2pc}  (k,a)\ne(q,b)\end{cases}\label{smatrix}
\end{multline}
This expression is not yet a solution for the scattering amplitudes, as it depends on the unknown eigenfunction $\phi_{k,a}$ in the region where the defect is placed.
However,  we can already draw some general conclusions.

Indeed, the scattering amplitudes are not arbitrary functions, but they must satisfy the constraints Eq. \eqref{constr1} and Eq. \eqref{constr2}. These relations can be translated in \emph{sum rules} that the scattering amplitudes must obey and are enough to determine the structure of the local GGE. 

Further details are left for Appendix \ref{appScat}, while hereafter we simply report the result. A local observable locally relaxes to a ray-dependent GGE: it is convenient to consider the rays in the moving reference frame, which in laboratory coordinates are simply defined as $\zeta=j/t-v$. The ray-dependent excitation density $\eta_\zeta(k)$, which ultimately uniquely identifies the GGE through Eq. \eqref{GGEmatrix}, can be written as
\begin{multline}
\eta_{\zeta}(k)=\Big[\theta(-\zeta v^1(k))+\theta(\zeta v^1(k))\theta(|\zeta|-|v^1(k)|)\Big]\eta(k)\\
+\theta(\zeta v^1(k))\theta(|v^1(k)|-|\zeta|)\eta_\text{scat}(k)\,.\label{LQSS}
\end{multline}
where $\eta(k)$ is the initial excitation density, introduced in Section \ref{secGenIsing}. The function $\eta_\text{scat}(k)$ is the excitation density propagating from the defect and is defined as
\begin{multline}
\eta_\text{scat}(k)=\sum_{n=-\infty}^\infty\sum_{a=1,2}\int_{-\pi}^\pi dq\,\Bigg[ \delta(E^a(q)-E^1(k+2\pi n))\times\\
|v^1(k+2\pi n)||S_{1,a}(k+2\pi n,q)|^2\eta^a(q)\Bigg]\, . \label{scatdens}
\end{multline}
with the convention $\eta^1(q)=\eta(q)$ and $\eta^2(q)=1-\eta(q)$. 
Despite deriving Eq. \eqref{LQSS} and \eqref{scatdens} requires some rather technical steps, their meaning is readily explained on a physical ground.

At fixed momentum $k$, Eq. \eqref{LQSS} simply describes the progressive replacement of the prequench excitation density with the one emerging from the scattering event. Indeed, let's suppose that $k$ is such that $v^1(k)>0$ and fix a large time $t$. Then, all the points on the left of the defect ($\zeta t < 0$) have not yet scattered on the defect; similarly, the points $\zeta t > v^1(k) t$ have not yet been reached by the excitations emitted by the scattering on the defect. Therefore, 
the excitation density is unaffected and $\eta_\zeta(k) = \eta(k)$ in these regions. Instead, for $0<\zeta<v^1(k)$, the excitation density is replaced by the one produced by the scattering events, i.e $\eta_\zeta(k) = \eta_\text{scat}(k)$.

As it is clear from Eq. \eqref{scatdens}, $\eta_\text{scat}(k)$ receives contributions from excitations ($a=1$, fact associated with $\eta(q)$) and holes ($a=2$, associated with $1-\eta(q)$), weighted with the proper squared scattering amplitudes.
Finally, the sum over the integers in  Eq. \eqref{scatdens} accounts for the folding of momenta in the first Brillouin zone while coming back from the continuous to the discrete model. 
Overall, we have framed the quantum problem in a simple scattering picture, being the details of the defect completely encoded in the scattering amplitudes.

\subsection{General features of the LQSS}
\label{Secgen}

Equations \eqref{LQSS} and \eqref{scatdens} fully characterize the emergent LQSS, encoding the details of the defect only in the scattering amplitudes. 
Even though, the specific values of $S_{a,b}$ are eventually needed in order for Eq. \eqref{scatdens} to be predictive (see Section \ref{deltaSolution} and \ref{secSemiclassic} ), we can nevertheless sketch the qualitative behavior of the LQSS based only on kinematics. Thereafter, we discuss how the scattering features change for defect's velocities ranging from $v=0$ up to $v>v_M$.

Consider at first a static defect ($v=0$): in this case the complicated detour of mapping the discrete model in the continuous one is clearly superfluous and we must recover the simple solution we would have obtained from a direct analysis of the lattice problem.
This is readily understood looking at the scattering processes in \eqref{scatdens}, which are allowed by the energy conservation 
\begin{equation}
\label{enercons}
E^1(k_\text{\tiny IN}) = E^a(k_\text{\tiny OUT}) \;, \quad a = 1,2 \;.
\end{equation}
Note that, for the time being, we treat $k$ as ranging on the whole real axis and only at the end fold it to the Brillouin zone. A graphical representation of the possible scattering channels is reported in Fig.~\ref{figscatchannels} and hereafter explained.

For $v=0$, as it is clear from Eq. \eqref{scaten} we simply have $E^1(k)=\omega(k)$ and $E^2(k)=-\omega(k)$. This immediately prevents any solution in the form $E^1(k)=E^2(q)$: a hole can never be scattered into an excitation. Instead, the equation $E^1(k)=E^1(q)$ has infinitely many solutions because of the periodicity of $\omega$, but once they are folded back to the Brillouin zone they reduce to only two possibilities $\pm k$. Therefore, for a static defect, an excitation is either transmitted or reflected with opposite momentum as it should be.

As soon as we move to $v>0$, the scattering picture becomes more involved. In fact, the equation $E^1(k)=E^1(q)$ has now only a finite number (divergent in the limit $v\to 0$) of solutions. These zeros are no longer equivalent once folded back in the Brillouin zone. Furthermore, scattering of holes into excitations is also possible, since $E^1(k)=E^2(q)$ has non-trivial solutions.
The number of possible scattering channels decreases as $v$ grows until $v= v_M$: after this threshold, the equation $E^1(k)=E^1(q)$ possesses the unique solution $k=q$. 
This feature was already present in the XX model presented in Ref. \ocite{hopdef} and was at the root of the lack of a superluminal LQSS. However, in the Ising model the situation is different. In fact, the equation $E^1(k)=E^2(q)$ still has one (and only one) solution.
Thus, from the kinematical point of view, the scattering of a hole into an excitation is allowed: in Ref. \ocite{hopdef} such a transition was prevented by the already mentioned $U(1)$ symmetry, but in the generic case there is no reason obstructing such a scattering and a superluminal LQSS is expected (see again Fig. \ref{LQSSligthcone}).

Because of its peculiar mechanism of formation, a superluminal LQSS has clear differences from the subluminal counterpart.
They become manifest looking closely at the ray dependence encoded in Eq. \eqref{LQSS}: for $v<v_M$ the combination of Heaviside Theta functions provides a non trivial ray dependence to $\eta_\zeta$ (with a discontinuity in $\zeta=0$) which ultimately ensures a non trivial profile of local observables.
On the contrary, in the case $v>v_M$ we have $\min_k|v^1(k)|=|v-v_M|$, as it is clear from its very definition Eq. \eqref{vdef} and this ensures a $\zeta$-independent excitation density $\eta_\zeta(k)$ as long as $|\zeta|\le |v-v_M|$.
Considering a local observable $\mathcal{O}_j$ back in the laboratory frame, this guarantees the formation of a tail beyond the defect whose size grows as $t|v-v_M|$ where $\langle \mathcal{O}_j\rangle$ is translationally invariant. In the next section, we investigate explicitly the formation of such a plateau for a narrow defect and compare our predictions against numerics.

\section{An exact solution: the $\delta$-defect}
\label{deltaSolution}

\begin{figure*}[t!]
\includegraphics[width=0.4\textwidth]{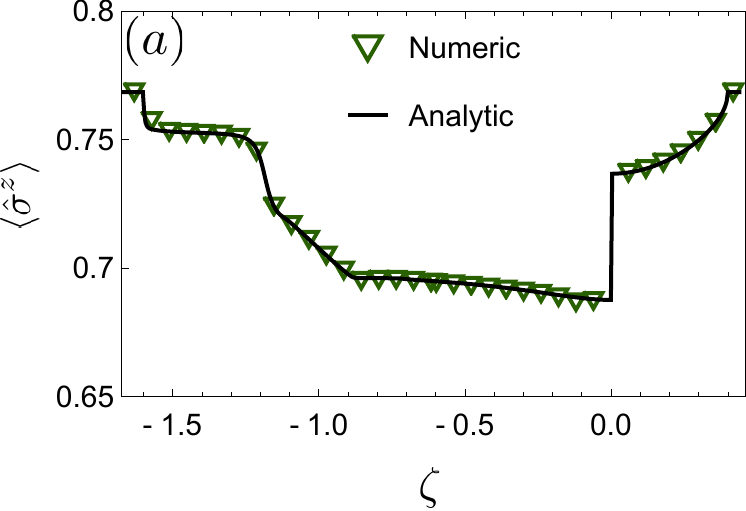}\hspace{2pc}
\includegraphics[width=0.4\textwidth]{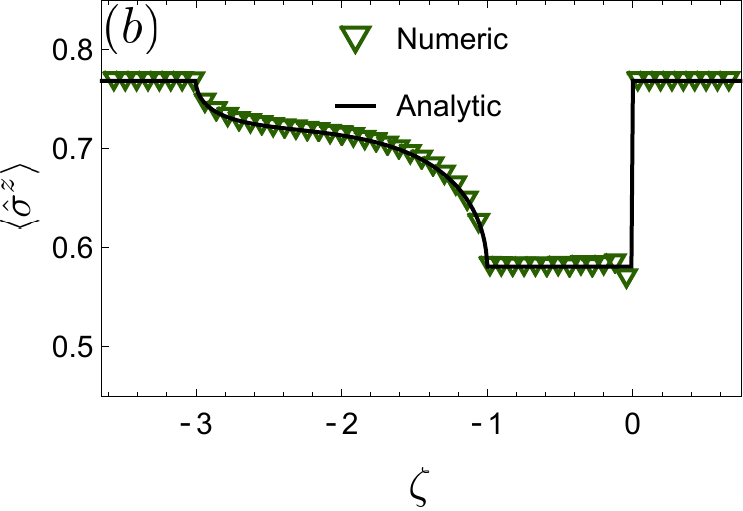}

\caption{\label{deltaLQSS} The analytic LQSS generated by a $\delta-$defect is compared with the numeric, in particular the local magnetization $\langle \hat{\sigma}^z\rangle$, trivially connected with the fermionic density $\hat{\sigma}_j^z=1-2\hat{d}^\dagger_j\hat{d}_j$, is considered in the subluminal (Subfigure $(a)$) and superluminal (Subfigure $(b)$) cases. Notice the presence of the plateau in the superluminal case, accordingly with the discussion of Section \ref{Secgen}. 
In both cases, the light velocity is $v_M=1$ and the ray $\zeta$ is measured with respect to the defect, which is thus located at $\zeta=0$.
Parameters Subfigure $(a)$: $h=1.15$, $c=0.6$, $v=0.6$, $1800$ sites.
Parameters Subfigure $(b)$: $h=1.15$, $c=2$, $v=2$, $1100$ sites. 
}
\end{figure*}

There exists one peculiar defect for which the scattering data Eq.~\eqref{smatrix} can be computed exactly and explicitly, i.e. the limit of an extremely narrow impurity described by a $\delta-$function
\be
\delta h(x)= c\delta(x)\, .\label{delta_def}
\ee
Before presenting the derivation, some additional comments are due: in Eq. \eqref{delta_def} the $\delta-$function is a genuine \emph{Dirac delta function}. Of course, as long as a static defect on a lattice is considered, Eq. \eqref{delta_def} is not well-defined: while pinching a lattice site gives a singularity, placing the delta in between two lattice sites will leave the system completely unaffected. However, the inconsistency disappears as soon as a moving defect is considered: for $v\ne 0$, Eq. (\ref{delta_def}) describes a ``kick" traveling along the lattice, impulsively acting on the system whenever the $\delta-$support meets a lattice site. As it should now be clear, this peculiar form of defect does not have a continuous limit $v\to 0$, as it happens instead for a smooth function $\delta h (x) $.  This example was also considered in Ref. \ocite{hopdef} for the XX spin chain and we can borrow the same techniques, even though the present case requires some extra technical refinements.

In Fig. \ref{deltaLQSS}, the forthcoming analytical results for a $\delta-$like impurity are compared with the numeric, finding excellent agreement. In particular, we show the  magnetization along the $z$ direction in the original spin model, directly associated with the fermionic density $\hat{\sigma}_j^z=1-2\hat{d}^\dagger_j\hat{d}_j$. It is worth stressing the presence of the already anticipated plateau in the superluminal case.

We now present the exact solution. If we directly approach Eq. (\ref{lippman}), formally replacing $\delta h(x)=c\delta(x)$ we would naively reach
\be
\label{phiwrong}
\phi_{k,a}(x)\stackrel{?}{=}\frac{e^{ikx}}{\sqrt{2\pi}}u_{k,a}+\mathcal{K}(x)c\sigma_z \phi_{k,a}(0)\, ,
\ee
where we pose $\mathcal{K}(x-x')=[E-H_0+i0^+]^{-1}(x,x')$. Despite the appealing simplicity, the above equation is \emph{not correct}. In the replacement of the $\delta-$function in the integral, it is implicit the continuity of the kernel, but this is not absolutely the case as $\mathcal{K}(0^+)\ne \mathcal{K}(0^-)$. This implies $\phi_{k,a}(0^+) \neq \phi_{k,a}(0^-)$ and it is therefore unclear how to properly interpret \eqref{phiwrong}. A careful way to proceed is to employ directly the Lippman-Schwinger equation
\be
\phi_{k,a}(x)=\frac{e^{ikx}}{\sqrt{2\pi}}u_{k,a}+\frac{1}{\sqrt{2\pi}}\mathcal{K}_{k,a}(x)W_{k,a}\, ,\label{lipW}
\ee
with $W_{k,a}$ a two dimensional vector coming from the (still) unknown regularization of the $\delta$ function. Eq. (\ref{lipW}) on its own is of course not a solution, being both $\phi_{k,a}$ and $W_{k,a}$ unknown, but we can provide an additional constraint from a direct analysis of the Schr\"oedinger equation (\ref{sch}). In fact, the singularity of the Dirac-$\delta$ must be balanced by the same singularity in the derivative of the wavefunction. Matching 
the most singular parts we deduce
\be
iv\partial_x\phi_{k,a}(x)+c\delta(x)\sigma^z\phi_{k,a}(x)=0\, , \quad x\to0\;. 
\ee
This is immediately translated in a jump discontinuity in $\phi_{k,a}$ around $x=0$
\be
\phi_{k,a}(0^+)=e^{i\frac{c}{v}\sigma^z}\phi_{k,a}(0^-) \, .\label{deltacond}
\ee

Combining Eq. \eqref{lipW} with the knowledge of the discontinuity, we can solve for $W_{k,a}$ in terms of the (still unknown) limits of the kernel $\mathcal{K}_{k,a}(0^{\pm})$.
\be\label{W_def}
W_{k,a}=\Big[ e^{i\frac{c}{v}\sigma^z}\mathcal{K}_{k,a}(0^-)-\mathcal{K}_{k,a}(0^+)\Big]^{-1}(1-e^{i\frac{c}{v}\sigma^z})u_{k,a}\, .
\ee
The scattering amplitudes are now accessible through Eq. (\ref{smatrix}) with the regularization
\be\int dx\, e^{-iqx}u^\dagger_{q,b}\sigma_z\delta h(x) \phi_{k,a}(x) \to  u^\dagger_{q,b}W_{k,a}\, .
\ee

The last ingredient we need is $\mathcal{K}_{k,a}(0^\pm)$, the latter being straightforwardly computable from its very definition. The calculations are somewhat lengthy but straightforward, following closely those presented in Ref. \ocite{hopdef}. Here we simply quote the result 
\begin{multline}\label{k_delta}
\mathcal{K}_{k,a}(0^\pm)=\pm\frac{i}{2v}+\sum_{b=1,2}\mathcal{P}\int \frac{dq}{2\pi} \frac{u_{q,b}\,u^\dagger_{q,b}}{E^a(k)-E^b(q)} \\
+\sum_{b=1,2}\int_{-\infty}^\infty dq\, \delta(E^a(k)-E^b(q))\frac{u_{q,b}\,u^\dagger_{q,b}}{2i}\, ,
\end{multline}
where $\mathcal{P}$ means the singular points must be regularized via principal values.

\section{The Semiclassical approximation}
\label{secSemiclassic}
Semiclassical methods are a very powerful tool, providing even accurate predictions for quantum systems in certain regimes of validity \ocite{wigners1,wigners2,wigners3,wigners4,wigners5,refA3}.
The semiclassical approach in describing a moving defect was already applied in Ref \ocite{hopdef} to the case of the XX model, but in that case no mayor differences between the quantum and classical case arose (even though purely quantum slow decaying corrections around the defect were observed, but still negligible in the scaling limit). 
Instead, in the case at hand a dramatic difference arises for superluminal defects where the semiclassical LQSS is absent. As discussed in Section \ref{Secgen}, superluminal LQSS are eventually due to the tunneling of a hole into an excitation, but such an event is classically forbidden and the defect becomes purely transmissive for $v>v_M$.

The derivation of the semiclassical approximation is left to Appendix \ref{derivation_semiclassical}, hereafter we simply report the result. 

In the hypothesis of weak inhomogeneity in space and time, the state is pointwise described by a GGE whose local momentum dependent excitation density $\eta(k,x)$ evolves through the classical \emph{Liouville Equation}

\begin{figure}[t!]
\includegraphics[width=0.45\textwidth]{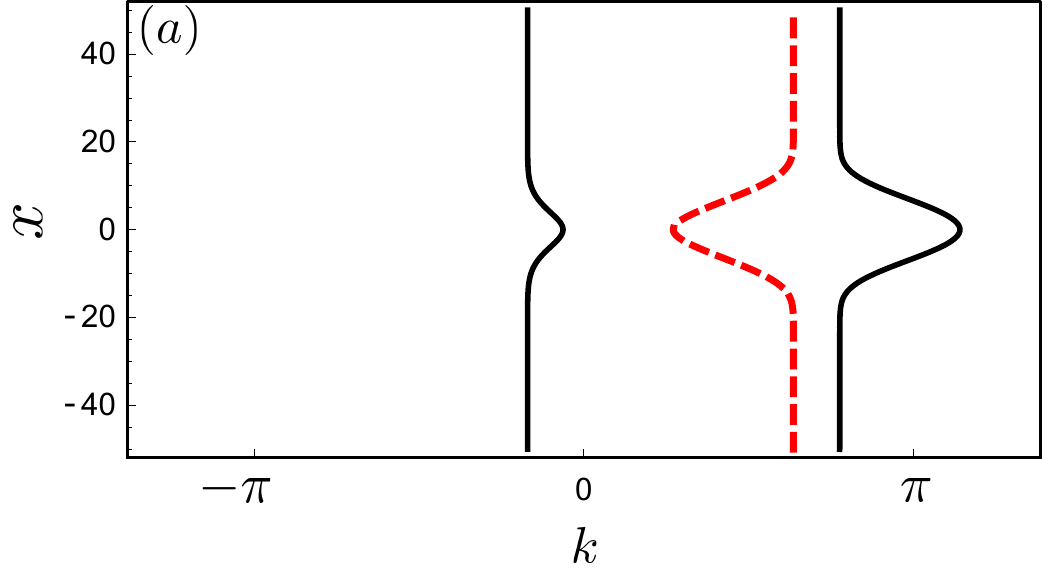}
\includegraphics[width=0.45\textwidth]{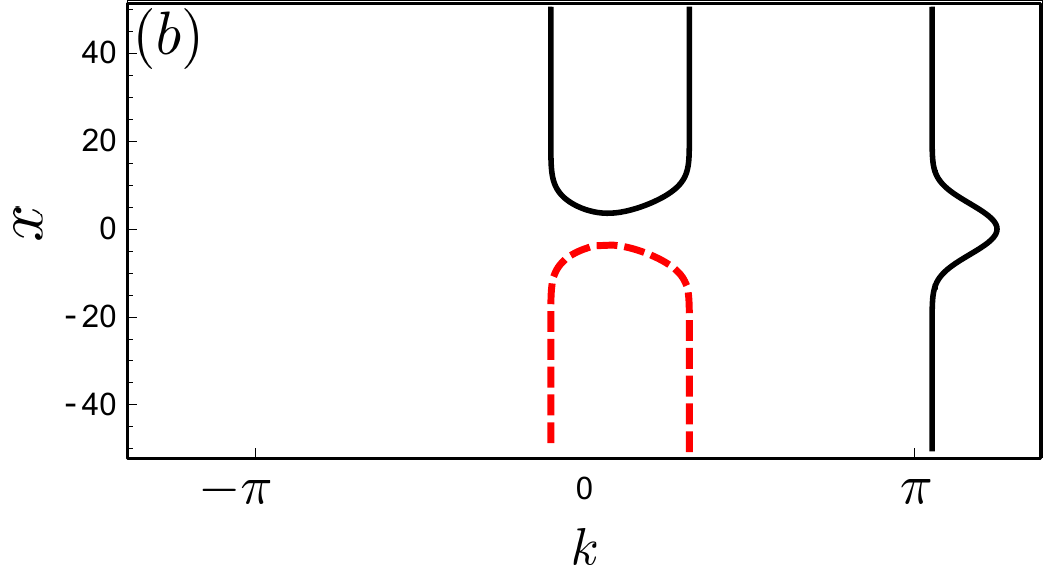}
\caption{
\label{energylevels}Energy levels $\omega(k,x)=\text{const.}$ for a transmitted (Subfigure $(a)$) and a reflected (Subfigure $(b)$) classical particle.
The classical scattering is fully determined by the energy levels: assuming the particle has initial momentum $k_\text{IN}$ infinitely far from the defect, the lines at constant level $\omega(k,x)=\omega(k_\text{IN},\infty)$ starting from $k_\text{IN}$ and $x=\pm \infty$ (depending on the direction of the incoming particle) determined wether the particle is transmitted or reflected (dashed red line).
Above, the case of a gaussian potential $V(x)=0.5 \,e^{-A x^2/2}$ moving at $v=0.5$ is considered, with $A=0.04 $ and constant magnetic field $h=1.3$. The initial momentum $k_\text{IN}$ has been taken $k_\text{IN}=2$ and $k_\text{IN}=1$ in Subfigure $(a)$ and $(b)$ respectively.}
\end{figure}

\begin{figure*}[t!]
\includegraphics[width=0.4\textwidth]{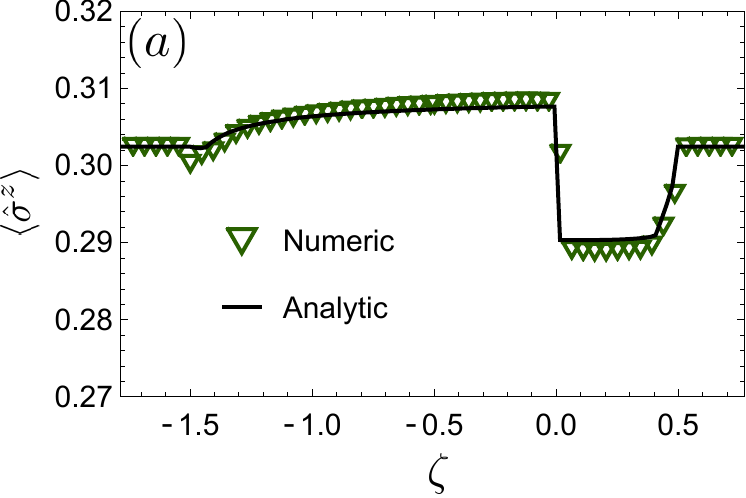}\hspace{2pc}
\includegraphics[width=0.4\textwidth]{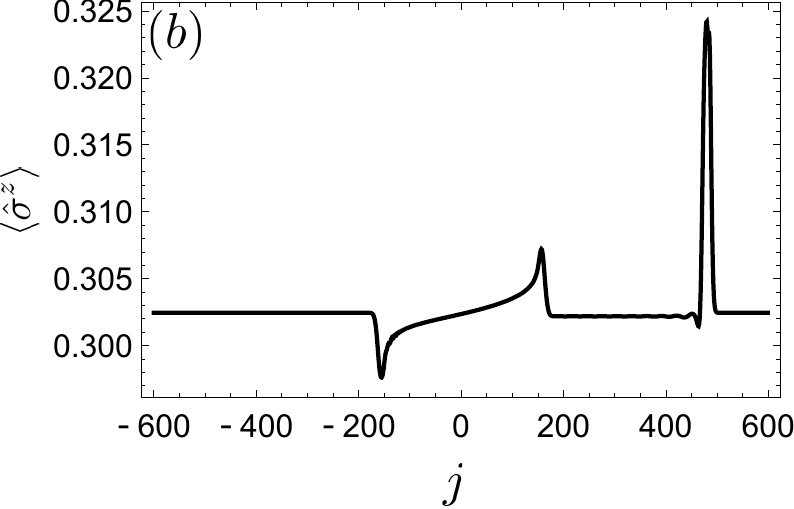}
\caption{\label{classicalLQSS}Subfigure $(a)$: The analytic semiclassical prediction for the LQSS is compared with the numerics, in the case of the local magnetization $\langle \hat{\sigma}^z\rangle$ and for a smooth defect.
Subfigure $(b)$: Within the semiclassical regime, a superluminal defect does not produce a LQSS. Above, the magnetization profile as a function of the lattice sites at a given time $t=150$. At late time, the system is affected only in the neighborhood of the defect (rightmost peak) where the scattering theory is no longer valid. At finite time, a transient included within the lightcone propagating from the initial position of the defect is displayed. Semiclassically, this transient is due to the quasiparticle initially sat on the defect that experience a sudden change of their energy and therefore of their velocity: a bunch of excitation starts traveling across the system leaving behind a hole. 
Parameters $(a)$:  $V(x)=0.5\, e^{-A x^2/2}$ ($A=0.04 $), $v=0.5$ and $h=1.3$. At time $t=0$ the system is initialized in a thermal ensemble with inverse temperature $\beta=0.5$, $1600$ sites are used.
Parameters $(b)$:  $V(x)= e^{-A x^2/2}$ ($A=0.04 $), $v=3$ and $h=1.3$. At time $t=0$ the system is initialized in a thermal ensemble with inverse temperature $\beta=0.5$, $800$ sites are used.
}
\end{figure*}

\be
\partial_t \eta(k,x)+\{\omega(k,x),\eta(k,x)\}_{k,x}=0\, ,\label{liouville}
\ee
where $\{\,\,\}_{k,x}$ are the Poisson's brackets and $\omega(k,x)$ acquires a parametric dependence through the inhomogeneous magnetic field
\be
\omega(k,x)=\sqrt{(\cos k-h(x))^2+\sin^2k}\, .
\ee

Therefore, $\eta(k,x)$ behaves as a classical density of particles evolving through the Hamiltonian $\omega$, i.e. obeying the equation of motion
\be
\dot{k}=-\partial_x\omega(k,x)\hspace{2pc} \dot{x}=\partial_k\omega(k,x)\, .\label{cleqm}
\ee

At the semiclassical level, the system is thus described by a classical gas of non interacting particles, moving in a background force where the defect is placed. In the classical picture the change of reference frame is most easily addressed, since in the semiclassical regime a coarse grain procedure is invoked and the lattice structure disappears (albeit it is still recognizable in the periodic $k-$dependence of $\omega(k,x)$).

In the moving reference frame, where the defect is at rest, the scattering amplitudes are determined by the classical equation of motion: a particle with momentum $k$ comes from infinity, scatters with the defect and then continues in a free motion with a new momentum. Being the classical equations fully deterministic, the scattering amplitude is non-zero only for a single scattering channel, determined by the equation of motion.

Notice that, being excitations and holes decoupled at a classical level, \emph{i)} if the system is initialized in the ground state, then no LQSS is produced and \emph{ii)} a superluminal defect does not create any LQSS. 

In the defect reference frame, the particles are governed by the following Hamiltonian
\be
\mathcal{H}_\text{cl}(k,x)=-vk+\omega(k,x)\, .
\ee

Even though classical, an exact solution of the equation of motion can be challenging, but in the perspective of solving the scattering problem only the trajectories in the phase space $(k,x)$, identified by the constant energy levels $\mathcal{H}_\text{cl}(k,x)=E$, are needed (see Figure \ref{energylevels}). For any incoming momentum $k_\text{IN}$ the proper outgoing momentum $k_\text{OUT}$ is then readily found.
The classical expression of the LQSS is then obtained simply replacing in Eq. (\ref{LQSS}) $\eta_\text{scat}$ with the excitation density of the incoming momentum, i.e. assuming $k_\text{IN}$ scatters in $k_\text{OUT}$ we simply have $\eta_\text{scat}(k_\text{OUT})=\eta(k_\text{IN})$.
The LQSS generated by  a smooth defect is tested with the semiclassical prediction in Figure \ref{classicalLQSS} Subfigure $(a)$, while the magnetization profile for a superluminal smooth defect is displayed in Figure \ref{classicalLQSS} Subfigure $(b)$.

\section{Friction }
\label{frictionsec}

\begin{figure*}[t!]
\includegraphics[width=0.4\textwidth]{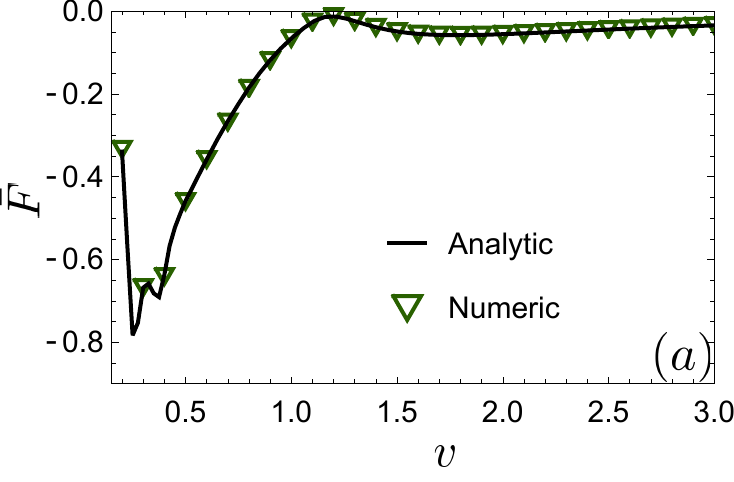}\hspace{2pc}
\includegraphics[width=0.4\textwidth]{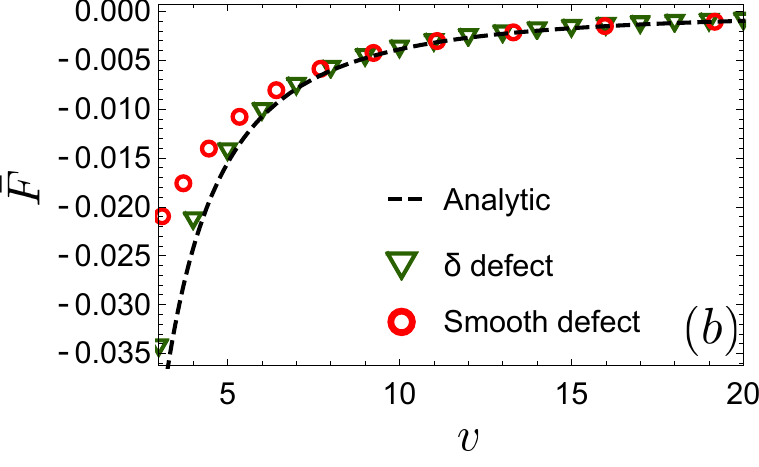}
\caption{\label{friction_fig}Subfigure $(a)$: The mean friction force $\bar{F}$ is plotted as a function of the defect's velocity $v$ for a $\delta-$like defect, showing excellent agreement between analytic and numerics. We considered velocities $v>0.2$, since the increasing number of singularities in Eq. \eqref{k_delta} makes difficult its practical evaluation for $v\to 0$. Parameters: $h=1.15$, $c=1$, the initial state is the ground state.
Subfigure $(b)$: The numerical mean friction force $\bar{F}$ for large velocities in the case of a $\delta-$like and a smooth defect, their normalization has been chosen in such a way the share the same asymptotic decay Eq. \eqref{Plargev} (dashed black line). Parameters: $h=1.15$, the strength of the $\delta-$defect is set $c=1$. The smooth defect is chosen as $\delta h(x)=0.56 \,e^{- 0.98\, x^2}$. The system has been initialized in the ground state.
}
\end{figure*}

In the previous sections we saw how the traveling defect affects the system as it moves, but the presence of the LQSS is associated to an energy cost.
This is particularly clear when the system is initialized in the ground state, where any non trivial effect is due to the creation of new excitations by the defect, which ultimately heats the system.
Reverting this point of view, we can equivalently say that the system exerts a friction force on the moving impurity.

The general issue of studying the emergence of friction forces on heavy impurities embedded in quantum thermodynamic baths has been extensively studied (see e.g. \ocite{rosch2010} for a review), with several experimental implications.
For example, in actual experiments heavy impurities are ultimately slowed down because of the recoil due to the scattering events (see e.g. Ref. \ocite{exp15}).

In our setup, the defect is considered as an external field and therefore it does not have a dynamics on its own, nevertheless this can be regarded as the zeroth order approximation of an heavy impurity, which is only slightly affected by each scattering event.
The instantaneous dissipated power is naturally defined as the derivative of the instantaneous energy of the system $P(t)=\partial_t\mathcal{E}(t)=\partial_t \langle \hat{H}(t) \rangle$ where $\hat{H}(t)$ is the full Hamiltonian of the Ising spins, including their interactions with the defect. From the dissipated power, we can readily define the friction force as $F=-P/v$.

However, the instantaneous dissipated power is affected by the microscopic dynamics around the defect, a region which lays beyond any scattering approximation. A more physical quantity which has a well-defined stationary regime is the time average of the dissipated power
\be
\bar{P}=-\bar{F}v=\lim_{\tau \to \infty}\frac{1}{\tau}\int_0^\tau dt\, P(t)=\lim_{\tau \to \infty}\frac{\mathcal{E}(\tau)-\mathcal{E}(0)}{\tau}\, \label{meanpower}
\ee
we can readily employ the results obtained so far. In fact, computing $\bar{P}$ only requires the total energy $\mathcal{E}(\tau)$ at a given large time and, in the $\tau\to \infty$ limit, any (finite) region around the defect can be entirely neglected. Therefore,  Eq. (\ref{meanpower}) can be computed simply looking at the system at large distances from the defect, where \emph{i)} the LQSS (\ref{LQSS}) is valid and \emph{ii)} the Hamiltonian reduces to the homogeneous one (\ref{Hising}).
In view of these considerations, computing $\mathcal{E}(\tau)$ amounts to locally consider the energy density using as excitation density that of the local LQSS, then integrate on the whole space.
This immediately leads to the simple expression 
\be
\bar{P}=\int d\zeta \int\frac{dk}{2\pi}\omega(k)[\eta_{\zeta} (k)-\eta(k)]\, \label{pw}.
\ee
Being $\eta_{\zeta}(k)=\eta(k)$ for $\zeta$ outside of the LQSS, the mean power is of course finite. 
Using the expression of $\eta_\zeta$ Eq. \eqref{LQSS}, the mean power is completely written in terms of the density of scattered momenta
\be
\bar{P}= \int\frac{dk}{2\pi}\omega(k)|v^1(k)|[\eta_\text{scat} (k)-\eta(k)]\, ,
\ee

which ultimately depends on the scattering data through Eq. \eqref{scatdens}.

In Fig. \ref{friction_fig}$(a)$ we compare the above analytic prediction against the numerics for a $\delta-$like defect, plotting $\bar{P}$ as a function of the defect velocity. Interestingly, in the limit of a very fast defect all possible impurities share the same behavior, decaying as $\sim 1/v$. This can be readily seen solving the Lippman-Schwinger equation \eqref{lippman} in the limit of large $v$ (see Appendix \ref{app_fastdef}), which ultimately leads to
\be\label{Plargev}
\bar{P}\simeq \frac{1}{v}\int\frac{dk}{2\pi}\omega(k)\left|u_{k,1}^\dagger\int dx' \delta h(x')\sigma_z u_{k,2}\right|^2(1-2\eta(k))\, .
\ee

This result must be commented in view of the semiclassical approximation of Section \ref{secSemiclassic}, which predicts exactly $\bar{P}=0$ as soon as the defect is superluminal, regardless its shape or the initial excitation density. However, there is no contradiction since the two results are valid in different regimes of approximation: as commented in Section \ref{secSemiclassic} and Appendix \ref{derivation_semiclassical}, the validity of the semiclassical approach requires weak inhomogeneities in space and time. Of course, if we consider a defect of given shape and strength and move it at higher and higher velocities, we ultimately leave the validity region of the semiclassical approximation, eventually ensuring the validity of Eq. \eqref{Plargev} (see Appendix \ref{app_fastdef} for the details). Actually, very fast defects are ultimately close to the $\delta-$potential, since they instantaneously kick each spin much faster than the speed of the signal between different sites $(v\gg v_M)$. In fact, all the defects at large velocities converge towards the same formula Eq. \eqref{Plargev}, which is ultimately the large velocity limit of a $\delta-$like defect, with effective strength $c=\int dx\,\delta h(x)$  (see  Fig. \ref{friction_fig}$(b)$).

An opposite, interesting limit is of course $v\to 0$. We consider the case of a $\delta-$like defect $\delta h(x)=c\delta(j-vt)$: physically, a moving defect in this form rotates a spin around the $z$ direction of an angle $c/v$ at intervals $\Delta t=1/v$.
While a direct analytical or numerical evaluation of our formulas becomes cumbersome because of the presence of infinitely many scattering channels, we can forecast the $v\to0$ limit of $\bar{F}$ based on simple physical arguments. As a matter of fact, as $v\to 0$, an infinite amount of time passes between two consecutive spin flips and the system manages to relax back to the initial state. Thus, in first approximation, the dissipated power is simply $\bar{P}=-\Delta \mathcal{E} v$, with $\Delta \mathcal{E}$ the energy needed to rotate one spin in the initial state. Therefore, as $v\to 0$, we get $\bar{F}=-\Delta \mathcal{E}$, which is readily computed as
\begin{multline}
\Delta\mathcal{E}=\langle e^{-i \frac{c}{2v}\sigma^z_j} \hat{H} e^{i\frac{c}{2v}\sigma^z_j}\rangle-\langle \hat{H}\rangle= \\
\frac{1}{2}\big\langle\left[(1-\cos(c/v))\sigma_j^x+\sin(c/v)\sigma_j^y\right](\sigma_{j+1}^x+\sigma_{j-1}^x) \big\rangle
\end{multline}
Above, $\hat{H}$ is the Ising Hamiltonian in absence of the defect \eqref{Hising} and the expectation value must be taken on the initial translational invariant state, thus the expression is independent on the actual choice of $j$.
From the above, we clearly notice the non analyticity of the $v\to 0$ limit if $c$ is kept constant, as we could have guessed from Eq. \eqref{W_def}. In particular, $\bar{F}$ becomes zero each time $c/v= 0(\text{mod}2\pi)$. A more sensible limit consists in sending $v\to 0$, but keeping $c/v$ fixed: in this case, the rotation angle is kept constant as $v\to 0$ and the force tends to a constant value.

We end this section commenting on the non-monotonicity of the friction force. It is tempting to re-inject this force in the equation of motion of the impurity itself, thus investigating the dynamics of the system coupled impurity. In particular, if an external force $\mathcal{F}$ is acting on the impurity, we would predict a drift velocity $v_d$ such that $\mathcal{F} + \overline{F}(v_d)=0$. 
Of course, this result would require further investigation, as it strongly relies on the possibility for the system to reach an LQSS, which contrasts with the impurity being a dynamical object. A comparison with the behavior obtained in similar settings (see for instance \ocite{rosch2010, gangardt2012}) following a different methodology could be useful to define the range of validity of our approach.

\section{Conclusions}
\label{secConclusions}

In this work we analyzed a local quench, consisting in the sudden activation of a localized defect moving at velocity $v$ in an otherwise homogeneous and thermodynamically large system.
While we considered the Ising model as a specific example, the same techniques can be immediately extended to \emph{arbitrary} quadratic models (fermionic or bosonic), or systems that can be mapped into this class, generalizing our previous work Ref. \ocite{hopdef}.

Our main focus has been the late-time features far from the defect, which we investigate by exploring the formation of a Local Quasi Stationary State. We provided \emph{exact} analytical results for an extremely narrow impurity and in the case of a smooth defect. Finally, we analyzed also the friction force exerted by the medium on the moving impurity. All our results have been tested against direct numerical simulations finding perfect agreement.

Moving impurities are the perfect benchmark to probe systems which possess a maximum velocity for the spreading of information, leading to the natural question whether or not the excitations of the system can cope with a fast moving defect. 
The Ising model manages to produce a LQSS even for superluminal defects, which nevertheless has rather characteristic features. Indeed, the superluminal LQSS is flat (i.e., ray-independent) beyond the defect up to a ray $v-v_M$ ($v_M$ the maximum excitation velocity). A part from the Ising model, the creation of a LQSS with the mentioned features is expected in all the free models, among which Ref. \ocite{hopdef} constitutes a remarkable exception due to the presence of additional symmetries. It remains open the difficult question of studying moving impurities in truly \emph{interacting} integrable models. Based on the experience gained so far, a superluminal LQSS is expected in general, even though the role of $U(1)$-symmetry in these cases would require further investigation. A fundamental difficulty is that, in most cases, defects break integrability: thus a scattering approach to connect the root densities on the two sides of the defect is only possible with some level of approximation.

A promising approach consists in considering smooth defects when Generalized Hydrodynamic \ocite{GHD1,GHD2} (see in particular Ref. \ocite{GHD3}) can be applied. However, in the free case GHD reduces to the semiclassical approach, where no superluminal LQSS is observed: in this perspective, it is unclear whether GHD has any chance to capture the physics of superluminal LQSS.

Finally, it is important to clarify how the behavior we observed is affected by perturbations which break integrability. In general, we do not expect ballistic transport, thus no LQSS (either subluminal or superluminal) is expected. However, a maximum velocity of information will still be present. This ingredient can be at the root of universal phenomena, which are our intention to investigate in the near future.

\section{Acknowledgment}

A.B. is grateful to Tomaz Prosen and Bruno Bertini for useful discussion on related topics and to Achim Rosch for sharing insights on friction forces of mobile impurities.

\appendix
\section{Scattering theory}
\label{appScat}

This Appendix is devoted to some technical details needed in the derivation of the LQSS formula Eq. (\ref{LQSS}). First of all, we enlist some useful symmetries. Using that in the fermionic field $\hat{\psi}_j$ the first component is the Hermitian conjugated of the second (and this is reflected on a symmetry of the Green function) leads us to the conclusion
\be
E^a(k)=-E^{\bar{a}}(-k),\,\,\,v^a(k)=v^{\bar{a}}(-k),\,\,\,\sigma^x u_{k,a}=(u_{-k,\bar{a}})^*
\ee
with $a = 1$ or $2$ and $\bar{a}=2$ or $1$ respectively. In the Ising case, this can be immediately checked by the very definitions, but it holds true for any quadratic fermionic model.
A similar symmetry holds for the wavefunctions $\phi_{k,a}$ which must ultimately obey
\be
\phi_{k,a}=\sigma^x (\phi_{-k,\bar{a}})^*\, 
\ee
and leads to an useful symmetry of the scattering amplitudes
\be
S_{b,a}(q,k)=\Big[S_{\bar{b},\bar{a}}(-q,-k)\Big]^*\,.
\ee

Other important constraints obeyed by the scattering amplitudes are certain \emph{sum rules} derived by the orthonormality and completeness of the eigenfunctions $\phi_{k,a}$.
The two sum rules involve a sum over the incoming and outgoing momenta respectively
\begin{widetext}

\be
\delta(k-k')\delta_{a,a'}=\sum_{b=1,2}\int dq\,\delta(E^a(k)-E^b(q))\delta(E^a(k')-E^b(q))[S_{a,b}(k,q)]^*S_{a',b}(k',q)|v^a(k)|^2\label{sumrule1}\, ,
\ee

\be
\delta(k-k')\delta_{a,a'}=\sum_{b=1,2}\int dq\,  \delta(E^a(k)-E^b(q))\delta(E^{a'}(k')-E^b(q)) [S_{b,a}(q,k)]^*S_{b,a'}(q,k')\left|v^b(q)\right|^2\label{sumrule2}\, .
\ee
\end{widetext}

The first sum rule can be derived from the completeness relation\footnote{Rather than considering directly the completeness relation (\ref{constr1}), is more convenient looking at it in the Fourier space.} (\ref{constr1}) , while the second follows from orthonormality (\ref{constr2}). 
Even though not difficult, their derivation requires some lengthy manipulations and will not be reported: the interested reader can refer to Ref. \ocite{hopdef} for similar identities in the closely related (but simpler) case of the XX chain.

With the help of the sum rules, the late-time behavior of the Green function can be extracted and from this the LQSS. Rather than considering the Green function in the coordinate space, it is more convenient to take a Fourier transform with respect to the second coordinate. In this way, the connection between the continuous and the lattice model is best displayed, as it is clear from Eq. (\ref{gfourier}). Applying the desired Fourier transform to the decomposition (\ref{Geigen}), we reduce ourselves to compute
\be
\mathcal{G}'_{x,k}=\sum_{a=1,2}\int \frac{dq}{2\pi} e^{-iE(q,a) t}\phi_{q,a}(x) \int dx'\, \phi^\dagger_{q,a}(x')e^{ikx'}\, \,\label{GF}
\ee

\begin{figure}[t!]
\advance\leftskip-1cm
\includegraphics[width=0.45\textwidth]{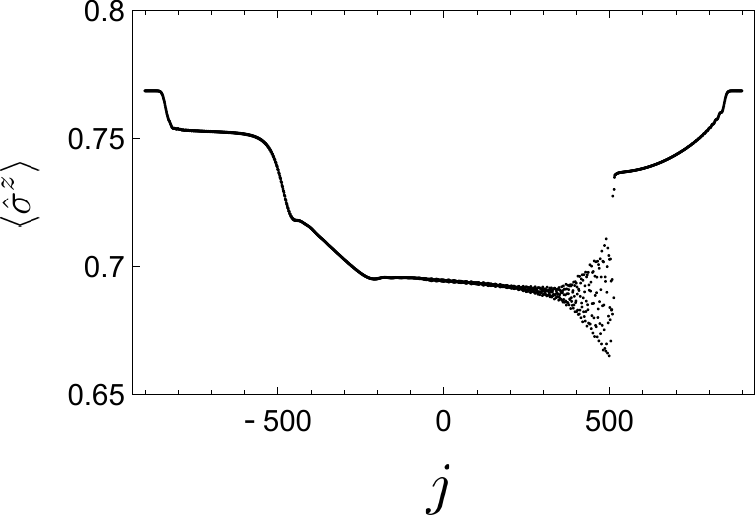}\\
\caption{\label{figosc} Oscillations of local observables can be present in proximity of the defect due to finite time/size effects. Above, we numerically compute the local magnetization $\langle\hat{\sigma}^z\rangle$. The same parameters of Figure \ref{deltaLQSS}  Subfigure $(a)$ have been used, but at finite time $t\simeq 850$. The defect, activated in $j=0$ at time $t=0$, is now placed nearby site $j \simeq 500$ where the discontinuity occurs.
}
\end{figure}
and we focus on the limit of large times and far from the defect. In this assumption we can \emph{i)} compute $\int dx'\, \phi^\dagger_{q,a}(x')e^{ikx'}$ replacing the wavefunction with its approximation in the scattering region and \emph{ii)} use the same approximation for $\phi_{q,a}(x)$ as well.
In computing the scalar product we face the Fourier transform of the Heaviside theta
\begin{widetext}
\be
\int dx'\, \phi^\dagger_{q,a}(x')e^{ikx'}\simeq \frac{1}{\sqrt{2\pi}}\Bigg(\frac{iu^\dagger_{q,a}}{\text{sgn}(-v^a(q))(k-q)+i0^+}+
\sum_{b=1,2}\int dq'\,\frac{i\delta(E^a(q)-E^b(q')) \theta(xv^b(q'))|v^b(q')| S^*_{b,a}(q',q)}{\text{sgn}(v^b(q'))(k-q')+i 0^+}u^\dagger_{q',b}\Bigg)\, ,
\ee
\end{widetext}
where ``$\text{sgn}$" is the sign function. This scalar product must now be used in (\ref{GF}): in the large time limit all the smooth contributions in $k$ vanish, only the singular parts of the above scalar product matters. These singularities, combined with the oscillating exponentials in $\phi_{q,a}(x)$, contribute again as Fourier transforms of Heaviside Theta functions leading to the final result (in the same notation of Eq. \eqref{gfourier})
\begin{widetext}
\begin{multline}
 \tilde{\mathcal{G}}_{x,k}(t)\simeq\sum_{a=1,2}\frac{e^{-iE^a(k) t}}{2\pi}\bigg[(\theta(-\zeta v^a(k))+\theta(\zeta v^a(k))\theta(|\zeta|-|v^a(k)|))e^{ikx} u_{k,a}+\\
\sum_{b=1,2}\int dq \delta(E^b(q)-E^a(k))\theta(|v^b(q)|-|\zeta|)\theta(\zeta v^b(q))|v^b(q)|S_{b,a}(q,k)e^{iqx}u_{q,b}\bigg]u^\dagger_{k,a}\, ,\label{asymgreen}
\end{multline}
\end{widetext}
valid in the large time limit, where $\zeta$ is the ray $x/t$ in the defect reference frame. The sum rule (\ref{sumrule1}) is crucial in the simplification of the squared scattering amplitudes that arise while evaluating Eq. \eqref{GF}. The asymptotic form of the Green function can now be used in the computation of the two point correlators and the large distance/time behavior extracted. It must be said that corrections to the scattering wavefunction (\ref{asymp}) decay exponentially fast away from the defect and the Green function (\ref{asymgreen}) is valid in the same regime, provided the long time limit has been taken. However, when the Green function is used in the two point correlators to derive the LQSS, a further long distance approximation must be invoked, in order to get rid of space-oscillating terms. These oscillations, coming from the interference of different scattering channels, can be quite slowly damped in the distance from the defect, nevertheless they vanishes in the LQSS limit (see Fig. \ref{figosc} and also Ref. \ocite{hopdef}).
With this caveat, recognizing in the two point correlator a local GGE with density root (\ref{LQSS}) is a matter of a long and tedious (but straightforward) computation, where the symmetries and the sum rules presented at the beginning of the Appendix play a fundamental role in simplifying many terms.

\section{Derivation of the Semiclassical Limit}
\label{derivation_semiclassical}

We derive the semiclassical limit directly in the more general case of an arbitrary free fermionic theory, where the fermionic field $\hat{\psi}_j$ is subjected to the linear equation of motion
\be
i\partial_t\hat{\psi}_j=\sum_{j'}\mathcal{H}_{j'}(L^{-1}j)\hat{\psi}_{j+j'}\, ,
\ee 
where the matrix $\mathcal{H}_{j'}(x)$ is a smooth function of its argument, the scale $L$ is introduced in order to make the notion of weak inhomogeneity limit precise, letting $L\to\infty$. Thereafter we consider the two point correlator $C$ 
\be
C(xL^{-1},s)=\langle \hat{\psi}_{x+s/2}\hat{\psi}^\dagger_{x-s/2}\rangle\, \,
\ee
which is a two by two matrix and we aim to follow its time evolution in the limit $L\to \infty$. Above, the coordinate $s$ has integer values and, in order to respect the underlying lattice, the coordinate $x$ should be (half)integer in order to ensure $x\pm s/2$ being a true lattice label (thus, an integer). However, in the limit $L\to \infty$ we will eventually face a coarse graining procedure and we can drop any specific requirement on $x$, considering it as if it was a continuous real variable.
Of course, in the limit of weak inhomogeneity, a non trivial time evolution needs longer times to be observed. Therefore, we need to describe the dynamics up to times $t=L\tau$, scaling with the regulator $L$. Three extra assumptions are needed on the two point correlator: \emph{i)} $C(y,s)$ is initially locally described by a GGE, \emph{ii)} $C(y,s)$ is smooth in the first argument and in the rescaled time $\tau$ and \emph{iii)} it decays faster than any power in $s$, i.e. $\lim_{|s|\to \infty} C(y,s) |s|^n=0$, for any $y,n$. These hypothesis are needed to justify the derivative expansion we are going to use: they must be satisfied by the initial condition and checked self consistently through the time evolution\footnote{The present derivation is not fully justified for slow decaying correlators as in the case of the ground state, since they violate the second hypothesis}.
The two point correlator satisfies the Heisenberg equation of motion
\begin{widetext}
\be
i\partial_t C(y,s)=\sum_j (\mathcal{H}_j(y+L^{-1}s/2,j)\otimes \text{Id})C(y+L^{-1}j/2,j+s)-\sum_j C(y+L^{-1}j/2,s-j)(\text{Id}\otimes\mathcal{H}_j(y-L^{-1}s/2,j) )\label{correq}
\ee
\end{widetext}
where with $\otimes$ we mean the tensor product.
This is nothing but an equivalent form of the Moyal equation used in Ref. \ocite{F17} to derive higher order corrections to hydrodynamics in the XX spin chain. In that case, an homogeneous post quench Hamiltonian was assumed.
For $L\to\infty$, we can Taylor expand the above equation and any derivative carries an extra $L^{-1}$ factor.
Rather than using the two point correlator in the coordinate space, it is convenient to partially transform it and use the Wigner representation \ocite{wigner1,wigner2}
\be
C(y,s)=\int_{-\pi}^\pi \frac{dk}{2\pi} e^{iks}\mathcal{C}(y,k)\, .\label{wigner}
\ee
And define 
\be
\Omega_k(y)=\sum_j e^{ikj}\tilde{\mathcal{H}}_{j}(y)\,.
\ee
In terms of these quantities, Eq. (\ref{correq}) up to $\mathcal{O}(L^{-2})$ corrections becomes (the explicit momentum/coordinate dependence is dropped for simplicity)
\begin{multline}
i\partial_t \mathcal{C}= \Omega\otimes\text{Id}\, \mathcal{C}- \mathcal{C}\, \text{Id}\otimes\Omega^\dagger+ \frac{i}{2L}\partial_y\Omega\otimes\text{Id}\, \partial_k\mathcal{C}+\label{eqmmomentum}\\
\frac{i}{2L} \partial_ k \mathcal{C}\, \text{Id}\otimes\partial_y\Omega^\dagger-\frac{i}{2L}\partial_k\Omega\otimes\text{Id}\, \partial_y\mathcal{C}-\frac{i}{2L} \partial_y \mathcal{C}\, \text{Id}\otimes\partial_k\Omega^\dagger\, .
\end{multline}
Thanks to our assumptions, the two point correlator is initially described by an inhomogeneous GGE, thus we pose
\begin{multline}
\mathcal{C}(y,k)=\sum_{a=1,2}\eta^a(y,k)u_{k,a}(y)u^\dagger_{k,a}(y)+\\
+\frac{\xi^a(k,y)}{L}u_{k,a}(y)u^\dagger_{k,\bar{a}}(y)\label{twoexp2}
\end{multline}
Where $u_{k,a}(y)$ are the local Bogoliubov orthonormal eigenvectors (always existing for a quadratic fermionic Hamiltonian), i.e. they are eigenvectors of $\Omega$, thus $\Omega_{k}(y)u_{k,a}(y)=\omega_{a}(k,y)u_{k,a}(y)$. Above, $\bar{a}=1(2)$ if $a=2(1)$. $\eta^l$ will be later interpreted as the excitation/hole density, $\xi^a(y,k)$ describes the off diagonal corrections.

Initially, is set $\xi^a=0$ by hypothesis, but we need to show (and we will) that it remains $\mathcal{O}(L^0)$ during the whole evolution. Even though we know the exact expressions for $u_{k,a}$ and $\omega_a$ for the Ising model, we can proceed in full generality. The Bogoliubov eigenvectors \eqref{eigenvectors} can always been parametrized in term of the Bogoliubov angle, being the details of the model completely encoded in the dependence of the latter from the momentum and the various couplings. However, this is enough to say $\partial_\mu u_{k,a}=i(\partial_\mu \, \gamma_k) u_{k,\bar{a}}$ where $\mu$ can be either the momentum or the position. Using this information, plugging (\ref{twoexp2}) in (\ref{eqmmomentum}) and projecting on the diagonal/off diagonal parts, we readily obtain the following two sets of equations
\be
\partial_t \eta_a =\frac{1}{L}\Big(\partial_y\omega_a\partial_k\eta^{a}-\partial_k\omega_a\partial_y\eta^a\Big)\, , \label{almostsemiclassical}
\ee
\be
i\partial_t \xi^a =\xi^a(\omega_a-\omega_{\bar{a}})+\mathcal{F}_a\, .\label{eq50}
\ee
Where $\mathcal{F}_a$ is a polynomial of $\omega,\gamma,\eta$ and their derivatives, but its explicit form does not matter in the forthcoming argument. Eq. (\ref{almostsemiclassical}), once we rescale the time $t=L\tau$, is in the Liouville form we where aiming for (Eq. (\ref{liouville})). 

In order to be consistent with our assumptions, we should ensure that $\xi^a$ remains of order $\mathcal{O}(L^0)$ up to times $t=L\tau$. Notice that (\ref{eq50}) is in the form of an oscillating term plus a $\xi-$independent driving: if the driving does not contain any resonant frequency, then $\xi$ remains bounded in time as we desire. Since $\mathcal{F}_a$ is a polynomial of smooth functions of the rescaled time $\tau=L^{-1}t$, the frequencies contained in $\mathcal{F}_a$ are vanishing as $L^{-1}$, therefore no resonance occurs as long as the system is gapped $\omega_a\ne \omega_{\bar{a}}$. The gapless case requires more attention, since from (\ref{eq50}) is not guaranteed that $\xi$ remains bounded. In this case we should remember that we are interested in the correlation rather than in the Wigner function (\ref{wigner}): integrating over the momentum the net effect is that a gapless equation (\ref{eq50}) makes larger the corrections to the semiclassical regime and replacing the current $\mathcal{O}(L^{-1})$ with a slower power law decay.

\section{The fast defect limit}
\label{app_fastdef}

This Appendix is devoted to study the scattering theory in the limit of fast defects $v\gg v_M$ at the leading order in $v^{-1}$, but for arbitrary potentials. 
In order to do that, we can start from the Lippman-Schwinger equation \eqref{lippman}, in particular we can consider the integration kernel which explicitly reads
\be
\left[\frac{1}{E^a(k)-H_0+i0^+}\right]_{x,x'}=\int \frac{dq}{2\pi}\sum_b\frac{u_{q,b}u^\dagger_{q,b}e^{iq(x-x')}}{E^a(k)-E^b(q)+i0^+}\, .
\ee
In the limit of very fast defects we can approximate $E^a(k)\simeq -vk$: this approximation is reliable in the limit $v\gg v_M$. In this assumption, the kernel is readily worked out
\be
\left[\frac{1}{E^a(k)-H_0+i0^+}\right]_{x,x'}\simeq\frac{e^{ik(x-x')}}{vi}\theta(x'-x)\, .
\ee
Employing this kernel in the Lippman-Schwinger equation we get the following approximation
\begin{multline}
\phi_{k,a}(x)\simeq\frac{e^{ikx}}{\sqrt{2\pi}}u_{k,a}
+\\ \int dx' \frac{e^{ik(x-x')}}{vi}\theta(x'-x)\delta h(x')\sigma_z \phi_{k,a}(x')\, .
\end{multline}

Of course, we do not really need to seek an exact solution: the leading order in $1/v$ is readily obtained by mean of the Born approximation, i.e. the wavefunction $\phi_{k,a}(x')$ in the integral is replaced with the unperturbed solution. Employing the Born approximation in the above equation and then reading the scattering amplitudes by mean of a direct comparison with Eq. \eqref{asymp}, we get at first order in $1/v$ the following excitation/hole scattering amplitude
\be
S_{\bar{a},a}(k,k)\simeq\frac{1}{vi}u_{k,\bar{a}}^\dagger\int dx' \delta h(x')\sigma_z u_{k,a}\, ,
\ee
where, as usual, $\bar{a}=1(2)$ is $a=2(1)$. Of course, in our approximation the energy conservation $E^{\bar{a}}(k')=E^a(k)$, which establishes the connection between ingoing and outgoing momenta, becomes trivial $k'=k$.
Of course, the momentum conservation has $\sim v^{-1}$ corrections $k'=k+\mathcal{O}(v^{-1})$, therefore the approximation is reliable if the defect cannot distinguish between momenta whose difference is $\sim v^{-1}$. This sets a bound on the typical width of the potential: let $\ell$ be the width of the potential, then our approximation is justified as long as $\ell v^{-1}\ll 1$.
Finally, the Born approximation is reliable in the small coupling limit, i.e. $v^{-1}\int dx\, \delta h(x)\ll 1$.

Aiming to obtain the density of the scattered momenta $\eta_\text{scat}(k)$ through Eq. \eqref{scatdens} we need the square of the scattering amplitude, which is of course $\mathcal{O}(v^{-2})$
\be
|S_{\bar{a},a}(k,k)|^2\simeq\frac{1}{v^2}\left|u_{k,\bar{a}}^\dagger\int dx' \delta h(x')\sigma_z u_{k,a}\right|^2\, .
\ee
For what it concerns the excitation/excitation and hole/hole squared scattering amplitudes, the Born approximation does not give us access to the $\mathcal{O}(v^{-2})$ order. However, a rather economic way to get $|S_{a,a}(k,k)|^2$ up to the desired order is to take advantage of the sum rules Eq. \eqref{sumrule1} and \eqref{sumrule2} which lead to
\be
|S_{a,a}(k,k)|^2\simeq1-\frac{1}{v^2}\left|u_{k,\bar{a}}^\dagger\int dx' \delta h(x')\sigma_z u_{k,a}\right|^2\, .
\ee
Plugging these scattering amplitudes in the definition of the scattered density Eq. \eqref{scatdens}, we finally get the desired leading order in $v^{-1}$
\be
\eta_\text{scat}(k)\simeq\eta(k)+\frac{1}{v^2}\left|u_{k,1}^\dagger\int dx' \delta h(x')\sigma_z u_{k,2}\right|^2(1-2\eta(k))\,  .
\ee

\section{Numerical methods}
\label{numApp}

In this Appendix we describe the numerical methods used to directly simulate the Ising chain with the moving impurity. We took advantage of the mapping in free fermions and directly solve the Heisenberg equation of motion Eq. (\ref{EQM}). A natural strategy would have been to write down the equation of motion for the two point correlator $\langle \hat{\psi}_j\hat{\psi}^\dagger_{j'}\rangle$ and then try to solve the time differential equation through a Runge Kutta method. However, we experienced huge instabilities in this method and thus adopt a transfer matrix approach hereafter explained.

The whole dynamics is entirely encoded in the Green function $G_{j,j'}(t)$ obeying the equation of motion (\ref{dG}): if we can compute $G_{j,j'}$, we can then reconstruct the correlation functions. For the seek of clarity, along this Appendix we slightly change the notation specifying two times in the Green function $G_{j,j'}(t_2,t_1)$, with the convention that it evolves the system from $t_1$ to $t_2$. Of course, we are ultimately interested in $G_{j,j'}(t,0)$.
The Green function of course respects the composition property 
\be
G_{j,j'}(t_3,t_1)=\sum_{l}G_{j,l}(t_3,t_2)G_{l,j'}(t_2,t_1)\, .\label{compG}
\ee
The presence of the moving defect surely breaks time translation, but we can notice that, after a time $1/v$ (where $v$ is the defect velocity), the defect is translated of one site. This implies that $G(t_2+nv^{-1},t_1+nv^{-1})$ and $G(t_2,t_1)$ ($n$ being integer) are simply related by a translation of the spatial indexes.
\be
G_{j,j'}(t_2+nv^{-1},t_1+nv^{-1})=G_{j-n,j'-n}(t_2,t_1)\, . \label{shG}
\ee
Combining Eq. (\ref{compG}) and (\ref{shG}) is an efficient way to quickly reach large times. In fact, assume we know $G_{j,j'}(v^{-1},0)$ (whose computation will be discussed soon), then we can compute $G_{j,j'}(2^n v^{-1},0)$ through the recursive relation
\begin{multline}
G_{j,j'}(2^n v^{-1},0)=\\
\sum_l G_{j-2^{n-1},l-2^{n-1}}(2^{n-1} v^{-1},0)G_{l,j'}(2^{n-1} v^{-1},0)\, .
\end{multline}
Even though each step in the recursive relation is not remarkably fast ($\sim N^3$), we can nevertheless reach large times exponentially fast exactly within machine precision. As a drawback, we cannot sample continuously the time evolution: even though suitable modifications of the above relation give us access to different times rather than $t=2^n v^{-1}$, the high cost of matrix products relegates us to a sparse sampling of the time evolution if large systems are considered.

We now discuss the last step, i.e. the computation of $G_{j,j'}(v^{-1},0)$. We will first address the case of the $\delta-$like impurity where the solution is exact up to machine precision, then we will describe how to approximate arbitrary well a smooth defect.
Concerning the $\delta-$like impurity, the evolution is homogeneous apart from those instants when the impurity travels through a lattice site. We assume at $t=0^+$ the impurity is right after the lattice site at $j=0$, then the evolution proceeds free until $t=v^{-1}$: we denote with $G^0_{j,j'}(v^{-1},0)$ the homogeneous Green function, exactly computable by means of an exact diagonalization of the homogeneous Ising model.
At time $t=v^{-1}$ the impurity suddenly kicks the system: the Green function has a discontinuity in time similarly to what the wavefunction (\ref{deltacond}) had in space and we get
\be
G_{j,j'}(v^{-1},0)=e^{-i\frac{c}{v}\delta_{j,1}\sigma_z}G^0_{j,j'}(v^{-1},0)\, ,
\ee
that is exact.
When an extended defect must be considered, we can approximate it with a sequence of $\delta-$ kicks. Given a smooth defect $\delta h(j-vt)$ we pose
\be
G_{j,j'}(n\Delta t,(n-1)\Delta t)=e^{-i\Delta t \sigma_z h(j-nv\Delta t)}G^0_{j,j'}(\Delta t,0)\, ,
\ee
with $\Delta t=v^{-1}/\mathcal{N}$, then $G_{j,j'}(v^{-1},0)$ is obtained composing together these Green functions. Increasing the number of steps used in the discretization $\mathcal{N}$, any smooth potential is approximated arbitrary well.

%%%%%%%%%%%%%%%%%%%%%%%%%%%%%%%%%%%%%%%%%%%%%%%%%%

\end{document}